\documentclass[12pt]{article}

\usepackage[top=1in,bottom=1in,left=1.25in,right=1.25in]{geometry}
\usepackage[T1]{fontenc}
\usepackage[utf8]{inputenc}
\usepackage{lmodern}
\usepackage{microtype}
\usepackage{amsmath,amssymb,bm}
\usepackage{booktabs,multirow,graphicx}
\usepackage{float}
\usepackage{caption}
\usepackage[dvipsnames]{xcolor}
\usepackage[colorlinks=true,
            linkcolor=NavyBlue,
            citecolor=NavyBlue,
            urlcolor=NavyBlue]{hyperref}
\usepackage{url}
\usepackage[numbers,sort&compress]{natbib}
\newcommand{\erank}{\mathrm{erank}}
\newcommand{\dout}{d_{\mathrm{out}}}
\newcommand{\Rbb}{\mathbb{R}}

\title{%
  \textbf{A Matched Spectral Benchmark of Quantum Inspired Feature Maps}%
}

\author{%
  Toheeb Ogunade$^{1}$,\quad Taofeek Kassim$^{2}$,\quad Etinosa Osaro$^{3}$\\[0.5em]
  \small$^{1}$Department of Computer Science, University of Lagos, Lagos, Nigeria\\
  \small$^{2}$Department of Physics, University of Lagos, Lagos, Nigeria\\
  \small$^{3}$Department of Chemical Engineering, University of Notre Dame, Indiana, USA\\[0.3em]
  \small\texttt{240805099@live.unilag.edu.ng},\quad
  \texttt{230808075@live.unilag.edu.ng},\quad
  \texttt{eosaro@nd.edu}
}

\date{}

\begin{document}

\maketitle

\begin{abstract}

Quantum machine learning is often motivated by the idea that quantum systems can
expose useful high-dimensional structure that is difficult to access with
classical models. We isolate one central component of this claim: the fixed
data-encoding map. Amplitude, angle, and basis encoding are evaluated as
deterministic feature maps for classical supervised learning under matched
output dimensionality and strong classical controls. The benchmark compares
these encodings against raw linear models, random Fourier features, polynomial
features, PCA, RBF SVMs, and shallow neural networks across diverse classical
datasets.

Rather than treating performance as a single endpoint, we analyze the geometry
of each representation through effective rank, condition number, centered kernel
alignment, predictive performance, and practical overhead. The resulting picture
is mechanistic: amplitude encoding can remove magnitude information through
unit-sphere normalization, angle encoding can become geometrically redundant
with raw linear features, and basis encoding can impose a binary Hamming
geometry that is poorly aligned with smooth decision structure. These findings
do not argue against quantum computation, however, they show that fixed quantum-inspired
encoding geometry alone is not a reliable source of machine-learning advantage
on classical data.

\end{abstract}

\section{Introduction}
\label{sec:intro}

Quantum machine learning rests on a powerful premise: quantum systems may
represent and manipulate information in feature spaces that are difficult to
access classically~\citep{havlicek2019,schuld2019}. Before any trainable
circuit, kernel evaluation, or measurement occurs, however, most quantum
learning pipelines perform a simpler operation. They encode classical data into
a quantum compatible representation. Amplitude, angle, and basis encoding are
the standard choices for this step. Each imposes a distinct geometry on the
input data. Amplitude encoding projects samples onto the unit sphere, angle
encoding maps features into trigonometric coordinates associated with
single qubit rotations, and basis encoding discretises continuous inputs into
binary strings.

This paper asks a narrow but important question: does the fixed encoding map
itself provide useful representation geometry? The question can be answered
without quantum hardware. If amplitude, angle, and basis encoding are
implemented as explicit classical feature maps, then their representational
value can be tested directly under matched dimensionality, matched data splits,
and strong classical controls. Such a benchmark does not decide whether quantum
computers can provide advantage. It decides whether one commonly invoked
mechanism for advantage, the geometry of the data loading map alone, survives a
controlled classical comparison.

The distinction matters. A quantum model may outperform a classical model
because of entanglement, measurement statistics, sampling complexity, trainable
circuit structure, or data generated by a quantum process. It may also appear to
outperform a weak baseline simply because the encoding expands the input
dimension. Without isolating the encoding step, these explanations are
confounded. A credible claim about quantum machine learning should identify
where the advantage enters. This benchmark removes one possible source of
ambiguity.

Existing work has shown that quantum kernels can be interpreted through
classical kernel theory~\citep{schuld2019}, that classical algorithms can
sometimes match proposed quantum speedups under appropriate data access
assumptions~\citep{tang2019}, and that classical models can match quantum
kernels on benchmark learning problems when given enough data or suitable
inductive bias~\citep{huang2021,kubler2021}. These studies motivate a careful
separation between the representation induced by a quantum feature map and the
broader computational model in which it is deployed. Accuracy alone does not
explain why a representation succeeds or fails. It also does not reveal whether
an apparent gain comes from useful geometry, increased dimensionality, numerical
conditioning, or a weak baseline.

This work evaluates amplitude, angle, and basis encoding as explicit,
deterministic feature maps in classical supervised learning. The benchmark spans
ten datasets across tabular, image, financial, physics, synthetic, and large
scale settings. Each encoding is compared against classical baselines under
matched output dimensionality, matched seeds, and comparable optimisation
effort. In addition to accuracy and macro F1, we analyse effective rank,
condition number, centered kernel alignment, timing, and memory usage.

The result is negative, but it is not merely a negative benchmark. Under matched
representational budget, fixed quantum inspired encodings do not produce a
statistically significant advantage over strong classical baselines on any
dataset tested. Their failures are also predictable. Amplitude encoding fails
when normalisation destroys class relevant magnitude structure. Angle encoding
is often nearly indistinguishable from the raw linear representation. Basis
encoding creates a different representation, but its Hamming geometry is poorly
aligned with the smooth decision structure of the benchmark tasks. These
findings sharpen the question of quantum advantage rather than dismiss it: the
encoding geometry alone is not enough.

The contributions of this paper are as follows:
\begin{enumerate}
  \item We present a matched benchmark of amplitude, angle, and basis encoding
        across ten datasets and five independent seeds, compared against strong
        classical baselines under controlled output dimensionality.
  \item We provide a spectral attribution analysis that links performance
        outcomes to effective rank, condition number, and kernel alignment.
  \item We identify three encoding specific failure modes: rank collapse for
        amplitude encoding, geometric redundancy for angle encoding, and Hamming
        geometry mismatch for basis encoding.
  \item We show that encoding overhead is not the limiting factor. The
        encodings are computationally cheap, but their induced representations
        are usually not competitive.
  \item We distinguish the scope of the result from claims about quantum
        hardware. The results rule out fixed encoding geometry as a standalone
        source of advantage on these classical datasets, but they do not rule
        out quantum advantage from trainable circuits, entanglement, sampling,
        or genuinely quantum structured data.
\end{enumerate}

\section{Related Work}
\label{sec:related}

\subsection{Quantum Feature Maps and Quantum Kernels}

Quantum feature maps provide a formal route from classical data to quantum
Hilbert spaces. A quantum circuit maps input data into a quantum state, and
classification can be performed through the induced kernel~\citep{havlicek2019}.
Variational quantum circuits can also be interpreted as kernel machines
operating in quantum feature spaces~\citep{schuld2019}. This connection makes
quantum learning models analyzable through representation geometry, kernel
alignment, and inductive bias. Data reuploading extends this idea by repeatedly
injecting classical inputs into parametrised circuits, increasing expressivity
even in small quantum systems~\citep{perez2020}.

\subsection{Classical Matching and Dequantisation}

Prior work cautions against attributing performance gains to quantum structure
without strong classical controls. A quantum inspired classical algorithm can
reproduce the speedup of a quantum recommendation algorithm under comparable
data access assumptions~\citep{tang2019}. In supervised learning, classical
models can match quantum models on benchmark datasets when sufficient training
data or suitable inductive bias is available~\citep{huang2021}. The performance
of a quantum kernel also depends strongly on the geometry of the data
distribution and the inductive bias induced by the kernel~\citep{kubler2021}.
These findings motivate baselines that control for dimensionality, model family,
and tuning effort.

\subsection{Limitations of Quantum Machine Learning Models}

Quantum learning models also have well documented failure modes. Variational
quantum algorithms can suffer from barren plateaus, where gradients vanish
rapidly with system size~\citep{cerezo2021}. Quantum neural networks may have
large effective dimension, but large capacity does not automatically imply
improved generalisation~\citep{abbas2021}. Quantum kernels can also
concentrate, making different samples increasingly similar and reducing
discriminative power in high dimensional regimes~\citep{thanasilp2024}.
Together, these results show that high dimensional quantum representations are
not inherently useful. Their geometry must match the task.

\subsection{Explicit Feature Maps in Classical Learning}

Classical machine learning has long used explicit feature maps to approximate
or replace kernels. Random Fourier features approximate shift invariant kernels
through random cosine projections~\citep{rahimi2007}. Polynomial expansions,
PCA projections, support vector machines, and neural networks provide additional
ways to reshape input geometry~\citep{scholkopf2002}. This literature makes the
central evaluation criterion clear: a feature map is useful only if the
similarity structure it induces is aligned with the target function.
Dimensional expansion alone is not enough.

\subsection{Positioning of This Study}

This work does not test quantum hardware and does not resolve the broader
question of quantum advantage in machine learning. It isolates one component
that appears in many quantum learning pipelines: the fixed encoding map. We
implement amplitude, angle, and basis encoding as deterministic feature
transformations and compare them against classical alternatives under matched
budget. The question is whether the encoding geometry itself is useful. In this
benchmark, the answer is no. The spectral analysis explains why.

\section{Methods}
\label{sec:methods}

\subsection{Problem Setting}
\label{sec:problem}

Let $\mathbf{X}\in\Rbb^{n\times d}$ denote a dataset with $n$ samples, $d$
input features, and labels $\mathbf{y}$. A fixed encoding is a deterministic map
\begin{equation}
    \phi: \Rbb^d \rightarrow \Rbb^{\dout},
\end{equation}
which transforms each sample before training a downstream classifier. We
evaluate whether the encoded representation $\phi(\mathbf{X})$ improves
predictive performance relative to classical feature maps of comparable output
dimension.

All encodings are fitted only on training data when training statistics are
required. The fitted transformation is then applied to the held out test set. No
quantum circuit simulator, quantum measurement routine, or quantum hardware
backend is used. The benchmark therefore measures the classical
representational effect of the encoding map itself.

\subsection{Quantum Inspired Encodings}
\label{sec:encodings}

\subsubsection{Amplitude Encoding}

For an input vector $\mathbf{x}\in\Rbb^d$, amplitude encoding produces
\begin{equation}
  \phi_{\mathrm{amp}}(\mathbf{x}) =
    \frac{\mathbf{x}}{\|\mathbf{x}\|_2 + \varepsilon},
  \label{eq:amplitude}
\end{equation}
followed by zero padding to the nearest power of two,
\begin{equation}
  \dout = 2^{\lceil \log_2 d \rceil}.
\end{equation}
The constant $\varepsilon=10^{-12}$ prevents division by zero. This
transformation mirrors the unit norm constraint of quantum state amplitudes.
Its central geometric consequence is that all samples are projected onto the
unit hypersphere, which removes absolute magnitude information.

\subsubsection{Angle Encoding}

Angle encoding maps each scaled feature to a pair of trigonometric coordinates
motivated by a single qubit rotation:
\begin{equation}
  R_y(\theta)|0\rangle =
  \cos(\theta/2)|0\rangle + \sin(\theta/2)|1\rangle.
\end{equation}
For classical input data, we implement this as
\begin{equation}
  \phi_{\mathrm{ang}}(\mathbf{x}) =
  [\cos(\theta_1/2),\sin(\theta_1/2),\ldots,
   \cos(\theta_d/2),\sin(\theta_d/2)],
  \quad \theta_i = \pi \tilde{x}_i,
  \label{eq:angle}
\end{equation}
where $\tilde{x}_i\in[-1,1]$ is obtained by min max scaling using training set
statistics. The output dimension is $\dout = 2d$. Each original feature
contributes one point on a unit circle.

\subsubsection{Basis Encoding}

Basis encoding discretises each continuous input feature into an $8$ bit
integer representation and concatenates the corresponding binary values:
\begin{equation}
  \phi_{\mathrm{bas}}(\mathbf{x}) =
    [b_1^{(1)},\ldots,b_8^{(1)},
     b_1^{(2)},\ldots,b_8^{(2)},\ldots]
    \in \{0,1\}^{8d}.
  \label{eq:basis}
\end{equation}
Here $b_k^{(i)}$ is the $k$th bit of the quantised value of feature $x_i$,
ordered from most significant to least significant bit. The output dimension is
$\dout=8d$. Unlike amplitude and angle encoding, basis encoding is not
differentiable because of the quantisation step.

\subsection{Classical Baselines}
\label{sec:baselines}

The benchmark includes classical alternatives designed to test whether QIE
performance can be explained by standard feature engineering, dimensional
expansion, or model capacity:
\begin{itemize}
  \item \textbf{Raw linear:} standardised input features followed by logistic
        regression.
  \item \textbf{Random Fourier features:} an explicit approximation to an RBF
        kernel with output dimension matched to the QIE representation
        ~\citep{rahimi2007}.
  \item \textbf{Polynomial features:} degree 2 or degree 3 polynomial
        expansions when computationally feasible.
  \item \textbf{PCA:} principal component projection with the same target
        dimensionality as the corresponding encoded representation, capped by
        the intrinsic rank of the data.
  \item \textbf{RBF SVM:} a radial basis function support vector machine with
        bandwidth tuned using the training set.
  \item \textbf{MLP baselines:} shallow neural networks implemented in
        scikit learn and PyTorch, included as adaptive nonlinear baselines.
\end{itemize}

For each dataset and seed, QIE methods and classical baselines are trained on
the same split. Comparisons are made against the best classical baseline
observed for that dataset. This is a conservative standard: a fixed encoding is
useful only if it can compete with the strongest classical alternative available
under the same evaluation protocol.

\subsection{Datasets}
\label{sec:datasets}

Table~\ref{tab:datasets} summarises the benchmark suite. The datasets span
small tabular classification, imbalanced financial classification, image
recognition, high energy physics, synthetic parity, synthetic high rank data,
and large scale multiclass classification. The externally sourced datasets are
Wine and Breast Cancer from the UCI Machine Learning Repository
~\citep{aeberhard1991wine,wolberg1995breast}, Dry Bean
~\citep{koklu2020drybean}, Credit Card Fraud
~\citep{dalpozzolo2015calibrating}, Fashion MNIST~\citep{xiao2017fashion},
CIFAR 10~\citep{krizhevsky2009cifar}, HIGGS~\citep{baldi2014higgs}, and
Covertype~\citep{blackard1998covertype}. The parity and high rank noise tasks
are synthetic controls generated by the benchmark code. The dataset roster was
fixed before analysing the results.

\begin{table}[tbp]
\centering
\caption{Benchmark dataset roster.}
\label{tab:datasets}
\small
\begin{tabular}{llrrrl}
\toprule
Dataset & Category & Samples & Features & Classes & Task \\
\midrule
UCI Wine           & Tabular     & 178       & 13      & 3  & multiclass \\
UCI Breast Cancer  & Tabular     & 569       & 30      & 2  & binary \\
Dry Bean           & Tabular     & 13{,}611  & 16      & 7  & multiclass \\
Credit Card Fraud  & Financial   & 284{,}807 & 30      & 2  & binary, imbalanced \\
Fashion MNIST      & Image       & 70{,}000  & 784     & 10 & multiclass \\
CIFAR 10           & Image       & 60{,}000  & 3{,}072 & 10 & multiclass \\
HIGGS              & Physics     & 500{,}000 & 28      & 2  & binary \\
High dimensional parity & Synthetic & 10{,}000 & 20      & 2  & binary, XOR \\
High rank noise    & Synthetic   & 5{,}000   & 200     & 2  & binary \\
Covertype          & Large scale & 581{,}012 & 54      & 7  & multiclass \\
\bottomrule
\end{tabular}
\end{table}

\subsection{Evaluation Protocol}
\label{sec:protocol}

All experiments are repeated over five independent seeds,
\begin{equation}
  \{7,42,99,1337,2026\}.
\end{equation} Dry Bean and high-dimensional parity were subsequently extended to ten seeds
  (adding $\{100,200,300,400,500\}$) after initial results showed large effect sizes
  that fell short of $\alpha=0.05$ under the five-seed design.
The same train test split is used for all methods within a given seed. The
primary metrics are accuracy and macro averaged F1. Macro F1 is included because
accuracy can be misleading on imbalanced datasets, especially Credit Card Fraud.

Statistical significance is assessed using paired tests over all available seeds.
For each QIE method and dataset, we compare the seed matched score against the
best classical baseline for that dataset. We report paired $t$ test results,
Wilcoxon signed rank tests, and Cohen's $d$ effect sizes. The nominal
significance threshold is $\alpha=0.05$.

To analyse representation geometry, we compute three quantities. First,
effective rank is defined as
\begin{equation}
\erank(\mathbf{X}) =
\exp\left(-\sum_i \hat{\sigma}_i \log \hat{\sigma}_i\right),
\label{eq:erank}
\end{equation}
where $\hat{\sigma}_i = \sigma_i / \sum_j \sigma_j$ are normalised singular
values. Second, the condition number is
\begin{equation}
\kappa(\mathbf{X}) = \frac{\sigma_{\max}}{\sigma_{\min}},
\end{equation}
with numerical safeguards for near zero singular values. Third, centered kernel
alignment measures similarity between Gram matrices induced by two
representations~\citep{kornblith2019}. CKA values near one indicate similar
geometry, while values near zero indicate distinct similarity structure.

\clearpage

\section{Results}
\label{sec:results}

\subsection{Predictive Performance}
\label{sec:perf}

Table~\ref{tab:accuracy} reports mean accuracy (up to ten seeds) for each QIE
method and the strongest classical baseline on each dataset. No QIE method is
consistently best. Angle encoding is usually the strongest QIE method on small
tabular datasets, but this does not translate into a significant win over
classical baselines. Amplitude encoding is highly unstable across datasets and
is competitive only on the high rank noise task. Basis encoding performs
moderately on several datasets but never surpasses the best classical model.

\begin{table}[tbp]
\centering
\caption{Mean accuracy (up to ten seeds) for QIE encodings and the best
classical baseline on each dataset. Bold indicates the highest value in the row.
The parity task is marked because all methods remain close to chance.}
\label{tab:accuracy}
\small
\begin{tabular}{lrrrr}
\toprule
Dataset & Amplitude & Angle & Basis & Best classical \\
\midrule
Wine                     & 0.628 & 0.972 & 0.917 & \textbf{0.978} raw linear \\
Breast Cancer            & 0.779 & \textbf{0.977} & 0.940 & 0.975 PyTorch MLP \\
Dry Bean                 & 0.260 & 0.928 & 0.920 & \textbf{0.931} PyTorch MLP \\
Credit Fraud             & 0.999 & 0.999 & 0.999 & \textbf{0.999} MLP \\
Fashion MNIST            & 0.821 & 0.823 & 0.819 & \textbf{0.859} PyTorch MLP \\
CIFAR 10                 & 0.378 & 0.306 & 0.328 & \textbf{0.457} RBF SVM \\
HIGGS                    & 0.643 & 0.658 & 0.669 & \textbf{0.748} PyTorch MLP \\
High dimensional parity   & 0.485 & 0.497 & 0.504 & \textbf{0.524} PyTorch MLP \\
High rank noise          & \textbf{0.795} & 0.780 & 0.700 & \textbf{0.795} RBF SVM \\
Covertype                & 0.587 & 0.725 & 0.724 & \textbf{0.859} MLP \\
\bottomrule
\end{tabular}
\end{table}

Across the full benchmark, QIE achieves zero statistically significant accuracy
wins against the best classical baseline. Of the 30 encoding-dataset comparisons, 27 are significantly worse at $p<0.05$
  and none are significantly better.
  The three non-significant cases are genuine negligible near-ties: angle on Wine
  ($d=-0.18$, $p=0.70$), angle on Breast Cancer ($d=+0.10$, $p=0.83$),
  and amplitude on high-rank noise ($d=-0.03$, $p=0.96$).
\subsection{Macro F1 and Imbalanced Classification}
\label{sec:f1}

Macro F1 supports the same conclusion as accuracy, but it is more revealing on
imbalanced data. On Credit Card Fraud, many methods achieve accuracy near 0.999
because the majority class dominates. Macro F1 separates the models more
clearly. The QIE methods achieve macro F1 values of 0.760, 0.845, and 0.887 for
amplitude, angle, and basis encoding, respectively, while the strongest neural
baselines reach approximately 0.914 to 0.915. Thus, even where QIE appears to
match accuracy, it does not match minority class performance.

\subsection{Statistical Analysis}
\label{sec:stats}

Effect sizes confirm that the benchmark outcome is not driven by isolated
outliers. Some QIE methods are only slightly below the classical baseline, such
as angle encoding on Wine ($d=-0.18$, $p=0.70$) and angle encoding on Breast
Cancer ($d=0.10$, $p=0.83$). Other failures are large and systematic. Amplitude
encoding on Dry Bean has an effect size of $d=-111.9$ with $p<0.0001$, and
amplitude encoding on Covertype has $d=-44.0$ with $p<0.0001$. These extreme
cases motivate the spectral analysis in Section~\ref{sec:analysis}, which shows
that amplitude encoding can collapse the usable rank of the representation.

\begin{figure}[h!]
  \centering
  \includegraphics[width=0.85\textwidth]{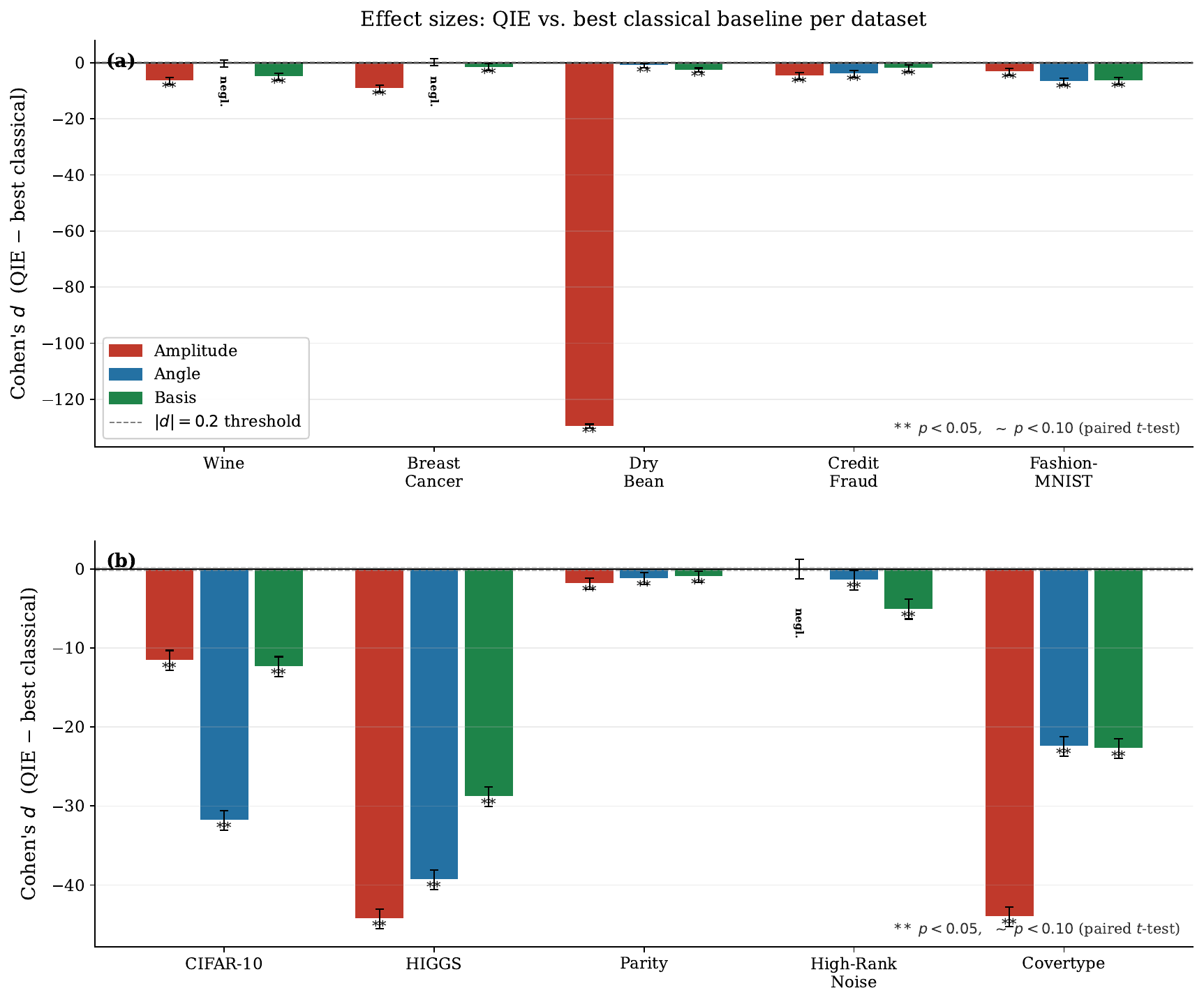}
  \caption{Forest plot of Cohen's $d$ effect sizes for all 30 QIE versus best
  classical comparisons. Negative values indicate that the QIE method is worse
  than the best classical baseline.   Error bars denote 95\% confidence intervals over five to ten seeds.}
  \label{fig:forest}
\end{figure}

\clearpage

\section{Spectral Attribution Analysis}
\label{sec:analysis}

Predictive performance establishes that fixed QIE maps do not outperform the
classical alternatives in this benchmark. The more important question is why.
We attribute the observed outcomes to three representation level mechanisms:
rank collapse, geometric redundancy, and geometry mismatch.

\subsection{Amplitude Encoding Collapses Rank When Magnitude Matters}
\label{sec:amplitude}

Amplitude encoding projects every sample onto the unit hypersphere. This is
faithful to the normalisation requirement of quantum state amplitudes, but it
can be harmful for classical data. If class information is partly encoded in
vector magnitude, then normalisation removes that signal before the classifier
is trained. The result is not merely a small rescaling effect. In several
datasets, the encoded matrix becomes nearly rank deficient.

\begin{table}[tbp]
\centering
\caption{Spectral properties of amplitude encoding on representative datasets.
Accuracy gap is measured relative to the best classical baseline.}
\label{tab:spectral_amplitude}
\small
\begin{tabular}{lrrr}
\toprule
Dataset & $\erank$ & $\log_{10}\kappa$ & Accuracy gap \\
\midrule
Dry Bean        &   1.04 & 9.76 & $-0.67$ \\
Wine            &   1.38 & 3.82 & $-0.35$ \\
Breast Cancer   &   1.64 & 5.88 & $-0.20$ \\
Covertype       &   3.24 & 5.73 & $-0.27$ \\
High rank noise & 197.3  & 0.35 & $-0.00$ \\
\bottomrule
\end{tabular}
\end{table}

Table~\ref{tab:spectral_amplitude} shows that the largest amplitude encoding
failures coincide with extremely low effective rank and poor conditioning. Dry
Bean is the clearest example: amplitude encoding reduces the representation to
an effective rank of 1.04 and produces a 67 percentage point accuracy deficit.
Wine, Breast Cancer, and Covertype show the same pattern at smaller scale.

The high rank noise dataset provides an important control. In this case, the
data are already distributed in a way that approximately fills the high
dimensional sphere. Normalisation therefore does not collapse the representation:
$\erank=197.3$ and $\kappa=2.22$. Under this condition, amplitude encoding
reaches near parity with the best classical method. This supports a falsifiable
interpretation: amplitude encoding is competitive only when the original data
geometry is already compatible with the unit sphere constraint.

\begin{figure}[h!]
  \centering
  \includegraphics[width=0.75\textwidth]{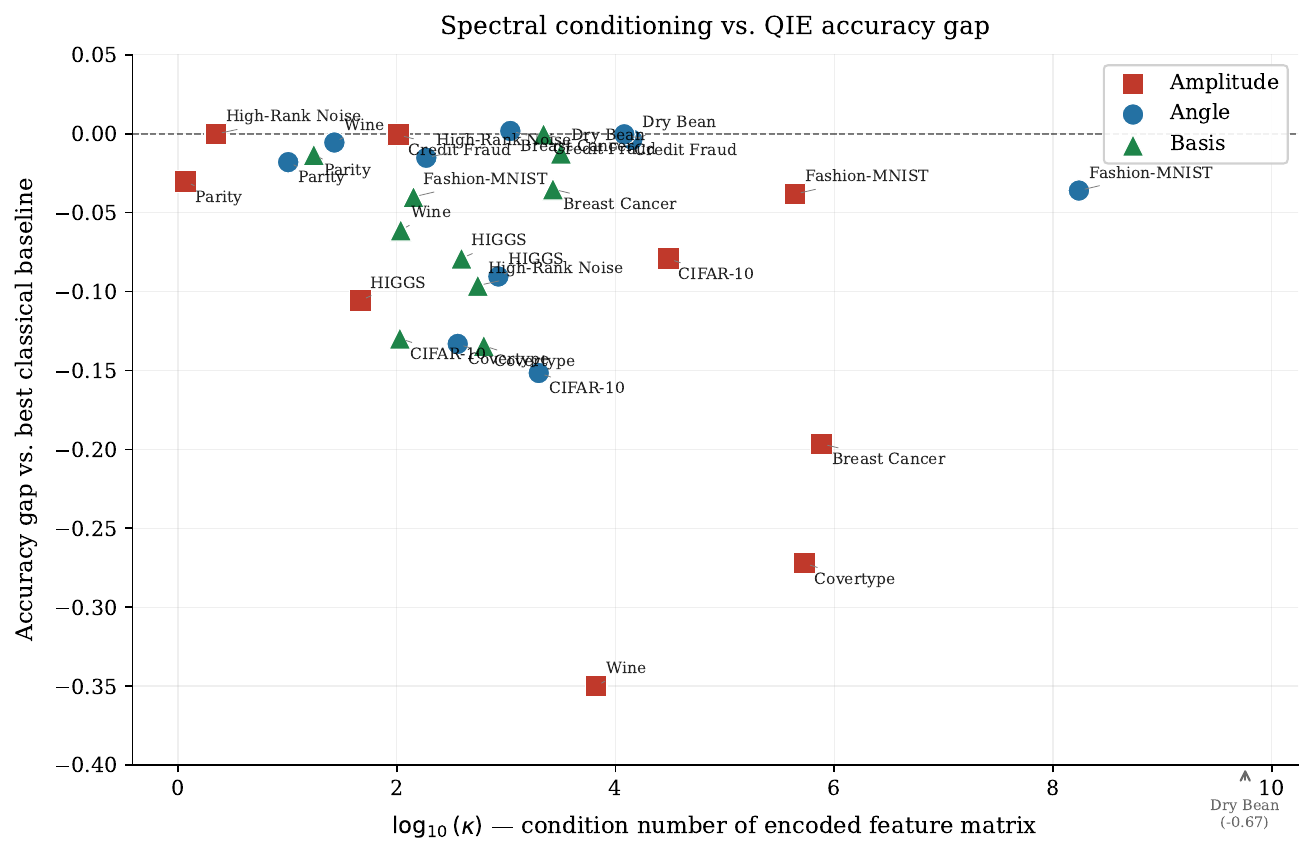}
  \caption{Condition number ($\log_{10}\kappa$) versus accuracy gap (vs.\ best
  classical baseline) for all three QIE encodings. Higher condition number is
  concentrated in amplitude encoding and correlates with larger performance
  deficits, consistent with the rank collapse failure mode.}
  \label{fig:kappa}
\end{figure}

\subsection{Angle Encoding Is Often Geometrically Redundant}
\label{sec:angle}

Angle encoding is the best performing QIE method on several datasets, but CKA
shows that this performance usually does not arise from a new representation.
Instead, the trigonometric map is often nearly aligned with the raw standardised
feature geometry under linear probing.

\begin{table}[tbp]
\centering
\caption{CKA between angle encoded features and raw standardised linear features.}
\label{tab:cka_angle}
\small
\begin{tabular}{lr}
\toprule
Dataset & CKA(angle, raw linear) \\
\midrule
High rank noise & 0.985 \\
Fashion MNIST   & 0.973 \\
Wine            & 0.971 \\
CIFAR 10        & 0.971 \\
High dimensional parity & 0.968 \\
Dry Bean        & 0.964 \\
Breast Cancer   & 0.957 \\
HIGGS           & 0.636 \\
Covertype       & 0.527 \\
Credit Fraud    & 0.514 \\
\bottomrule
\end{tabular}
\end{table}

On 7 of 10 datasets, $\mathrm{CKA}\geq0.95$. This indicates that angle encoding
induces a Gram matrix that is almost identical to the raw linear representation.
In these cases, strong performance should not be interpreted as evidence of a
quantum inspired representational advantage. It is better understood as a smooth
rescaling of the original features.

The three lower CKA cases, HIGGS, Covertype, and Credit Fraud, show that angle
encoding can become geometrically different from the raw representation.
However, it does not become better. This distinction is important: a feature map
can be novel without being useful. Angle encoding succeeds when it resembles a
classical linear representation and fails to improve when it departs from it.

\subsection{Basis Encoding Is Distinct but Misaligned}
\label{sec:basis}

Basis encoding has the opposite failure mode. It is often geometrically distinct
from the classical baselines, but that distinctness does not produce better
classification. By quantising continuous features into bit strings, the method
changes the metric structure of the data. Distances in the encoded space are
governed by bit flips rather than smooth changes in the original variables.

This binary Hamming geometry can be well conditioned and can have moderate
effective rank. However, the benchmark tasks largely contain smooth decision
structure. Images, tabular measurements, and physical observables usually change
continuously, while basis encoding introduces discontinuities at quantisation
thresholds. This mismatch explains why basis encoding can be spectrally healthy
and still underperform. It creates a different geometry, but not the right one
for these tasks.

\subsection{A Unified Spectral Interpretation}
\label{sec:prediction}

The three encodings fail for different reasons, but the outcomes are consistent
with one broader principle: a fixed feature map helps only when its induced
geometry matches the data distribution and the target function. Effective rank
and condition number capture whether the representation preserves useful
variation. CKA captures whether a representation is genuinely different from
classical alternatives. Predictive performance then reveals whether that
difference is useful.

\begin{figure}[h!]
  \centering
  \includegraphics[width=0.75\textwidth]{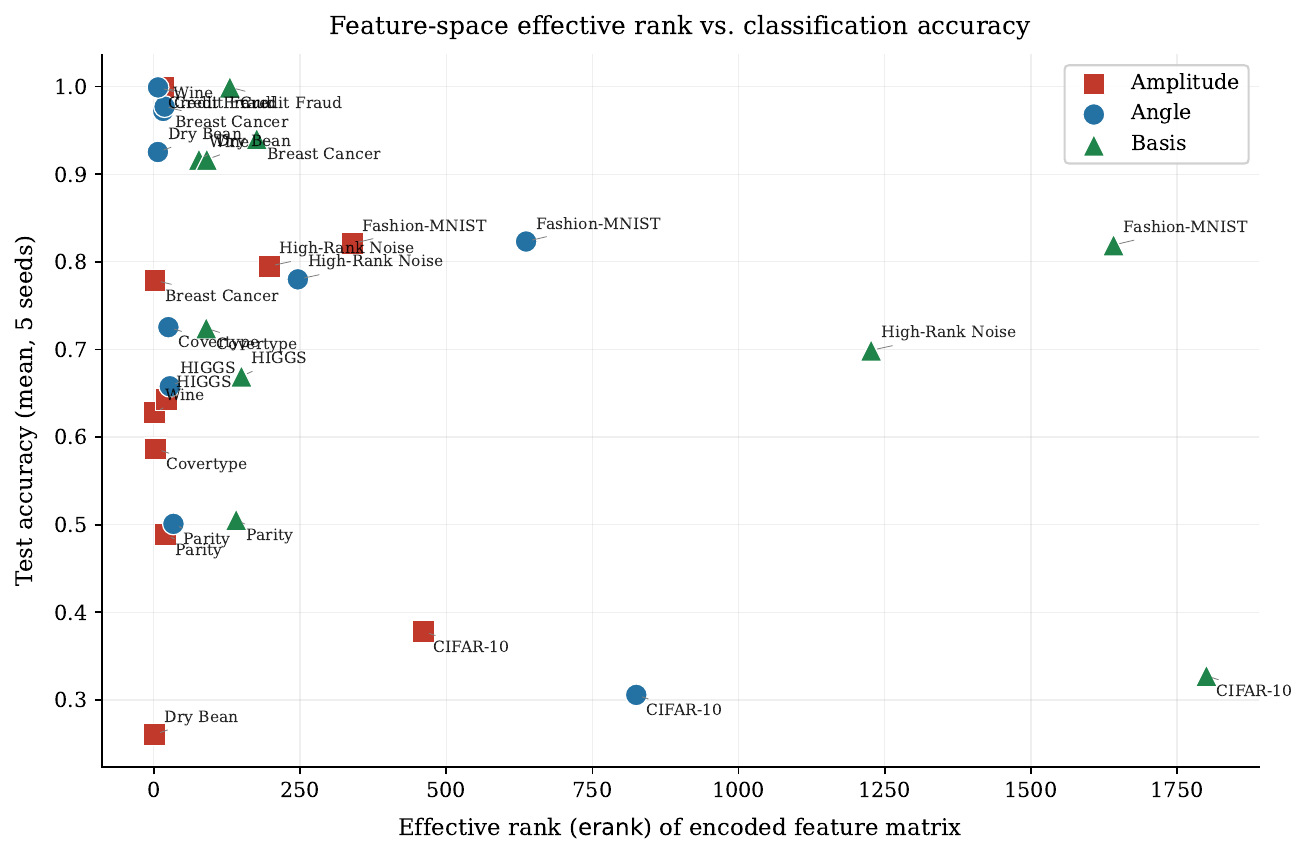}
  \caption{Normalised effective rank versus accuracy relative to a raw linear
  baseline across encodings and datasets. The near parity amplitude encoding case
  on high rank noise has $\erank/\dout=0.77$ and $\kappa=2.22$, consistent with
  the requirement that the output space be well filled.}
  \label{fig:erank}
\end{figure}

The high rank noise result is not an exception to the theory. It is the control
case that clarifies it. When the data already occupy the amplitude encoding
output space uniformly, the encoding does not destroy information and can match
classical performance. When the data do not satisfy this condition,
normalisation, redundancy, or binary mismatch dominate.

\clearpage

\section{Practical Overhead}
\label{sec:overhead}

A possible explanation for poor QIE performance is that the encodings may be too
expensive relative to their benefit. The overhead results do not support that
explanation. As shown in Table~\ref{tab:overhead}, the QIE maps are
computationally cheap. Amplitude encoding is essentially as fast as raw
standardisation, angle encoding remains below one second on average, and
basis encoding is still far cheaper than a degree 2 polynomial expansion.

\begin{table}[tbp]
\centering
\caption{Mean encoding overhead across the ten benchmark datasets.}
\label{tab:overhead}
\small
\begin{tabular}{lrrrr}
\toprule
Method & Enc. time (ms) & Train time (s) & Enc. mem. (MB) & Train mem. (MB) \\
\midrule
Amplitude QIE &     35.2 &    13.1 & 135.7 &   4.8 \\
Angle QIE     &    811.3 &   114.2 & 291.0 &   5.6 \\
Basis QIE     &  1{,}748 &    39.6 & 436.5 & 345.7 \\
\midrule
Raw linear    &     35.2 &    74.7 &  47.9 &  38.0 \\
RFF           &    193.0 &     2.1 &  65.2 &  24.5 \\
PCA           &  1{,}045 &    14.5 &  29.1 &  15.2 \\
Poly 2        & 10{,}981 &   144.9 & 818.5 & 178.4 \\
\bottomrule
\end{tabular}
\end{table}

This is a qualified positive result. If a dataset had geometry well matched to
one of these encodings, the transformation would be cheap enough to be useful in
practice. The limiting factor in this benchmark is not runtime. It is
representational alignment.

\clearpage

\section{Discussion}
\label{sec:discussion}

\subsection{What the Benchmark Shows}

Across ten datasets, fixed quantum inspired encodings do not provide a
statistically significant advantage over strong classical baselines. The result
is strongest when interpreted mechanistically rather than rhetorically.
Amplitude encoding fails when normalisation removes magnitude information and
collapses rank. Angle encoding is often competitive only because it closely
resembles the raw linear representation. Basis encoding is genuinely different,
but its binary geometry is not aligned with the tested tasks.

\subsection{What the Benchmark Does Not Show}

These results are not an argument against quantum computing. They do not test
quantum hardware, entanglement, quantum sampling, trainable variational circuits,
or classically hard quantum kernels. The study isolates the encoding step and
evaluates it as an explicit classical feature map. The correct conclusion is
therefore limited but important: fixed encoding geometry alone is not a
sufficient explanation for machine learning advantage on the classical datasets
considered here.

\subsection{Implications for Quantum Machine Learning}

The findings suggest that future claims about quantum machine learning should
separate the contribution of the data encoding from the contribution of the
trainable or quantum part of the model. If an observed advantage arises from a
fixed encoding, it should survive comparison against classical feature maps with
matched dimensionality and tuning effort. If it does not, then the advantage
must be attributed elsewhere. This separation can make positive claims stronger
by eliminating weaker explanations.

\subsection{Why Negative Results Matter}

A negative benchmark is useful when it narrows the space of plausible
mechanisms. In this case, the results show that dimensional expansion,
trigonometric lifting, and binary discretisation are not inherently beneficial
because they are quantum inspired. The geometry must match the problem. The high
rank noise control demonstrates this point directly: amplitude encoding can be
competitive when the data distribution satisfies the encoding's assumptions.

\subsection{Where Quantum Advantage May Still Appear}

The benchmark uses classical datasets. It is possible that quantum structured
data, such as molecular quantum states, Hamiltonian spectra, quantum simulation
outputs, or problems with provably hard classical representations, could behave
differently. It is also possible that learned quantum encodings or entangling
circuits could induce useful structure that fixed encodings do not. These are
not contradictions of the present result. They are precisely the settings where
future work should look if the fixed encoding map is not enough.

\subsection{Limitations}

The benchmark evaluates deterministic, parameter free encodings with linear
probes and shallow neural baselines. It does not include trainable quantum
circuits, data reuploading architectures, hardware noise models, or quantum
kernel estimation from finite measurement shots. The number of seeds is
sufficient for a controlled benchmark but still modest for very small datasets.
Finally, the dataset suite is broad but not exhaustive. The conclusions should
therefore be read as evidence against fixed QIE geometry as a general purpose
classical feature map, not as a universal statement about all quantum machine
learning models.

\section{Conclusion}
\label{sec:conclusion}

We benchmarked amplitude, angle, and basis encoding as explicit classical
feature maps under matched representational budget. Across ten datasets, up to ten
seeds, and 30 encoding dataset comparisons, none of the quantum inspired
encodings achieved a statistically significant accuracy advantage over the best
classical baseline. The failures were not arbitrary. Amplitude encoding often
collapsed rank by removing magnitude information. Angle encoding was usually
geometrically redundant with raw linear features. Basis encoding produced a
distinct but poorly aligned Hamming geometry. Encoding cost was small, so runtime
overhead was not the bottleneck.

The broader implication is that fixed quantum inspired encoding geometry, by
itself, is not a reliable source of machine learning advantage on classical
data. This does not diminish the possibility of quantum advantage from hardware,
entanglement, trainable circuits, or quantum structured data. Instead, it
clarifies what such an advantage would need to rely on. If quantum machine
learning succeeds, the source of the advantage is unlikely to be the fixed
encoding map alone.

\clearpage

\bibliography{references}

\clearpage
\appendix

\section{Supplementary Information}
\label{sec:supplementary}

\subsection{Reproducibility}
\label{app:repro}

All experiments use five fixed seeds, $\{7,42,99,1337,2026\}$. Dry Bean and
  high-dimensional parity were additionally run on seeds $\{100,200,300,400,500\}$
  to resolve underpowered comparisons identified in the initial analysis. Code,
configuration files, and per seed result files are available at
\url{https://github.com/qeinstein/qie_research}. Dataset preparation scripts are
included for externally sourced datasets. For a fixed seed and software
environment, the runner produces deterministic outputs; minor floating point
differences may occur across platforms.

\subsection{Extended Results}
\label{app:extended}

Tables~\ref{tab:ext1} and~\ref{tab:ext2} report mean accuracy and macro F1
across all methods and all datasets. QIE methods are listed first, followed by
classical baselines.

\begin{table}[H]
\centering
\caption{Mean accuracy and macro F1 for Wine, Breast Cancer, Dry Bean, and
Credit Card Fraud.}
\label{tab:ext1}
\resizebox{\textwidth}{!}{%
\begin{tabular}{lrrrrrrrr}
\toprule
 & \multicolumn{2}{c}{Wine} & \multicolumn{2}{c}{Breast Cancer}
 & \multicolumn{2}{c}{Dry Bean} & \multicolumn{2}{c}{Credit Fraud} \\
\cmidrule(lr){2-3}\cmidrule(lr){4-5}\cmidrule(lr){6-7}\cmidrule(lr){8-9}
Method & Acc & F1 & Acc & F1 & Acc & F1 & Acc & F1 \\
\midrule
Amplitude     & 0.628 & 0.502 & 0.779 & 0.710 & 0.260 & 0.059 & 0.999 & 0.760 \\
Angle         & 0.972 & 0.973 & 0.977 & 0.976 & 0.928 & 0.939 & 0.999 & 0.845 \\
Basis         & 0.917 & 0.919 & 0.940 & 0.936 & 0.920 & 0.931 & 0.999 & 0.887 \\
\midrule
Raw linear    & 0.978 & 0.978 & 0.974 & 0.972 & 0.925 & 0.936 & 0.999 & 0.854 \\
RFF           & 0.889 & 0.891 & 0.926 & 0.920 & 0.918 & 0.928 & 0.998 & 0.515 \\
PCA           & 0.928 & 0.929 & 0.951 & 0.947 & 0.908 & 0.919 & 0.999 & 0.852 \\
Poly 2        & 0.961 & 0.961 & 0.970 & 0.968 & 0.927 & 0.938 & 0.999 & 0.903 \\
Poly 3        & 0.950 & 0.950 & 0.958 & 0.955 & 0.929 & 0.940 & n/a   & n/a   \\
MLP           & 0.939 & 0.941 & 0.956 & 0.953 & 0.929 & 0.940 & 1.000 & 0.915 \\
RBF SVM       & 0.967 & 0.967 & 0.972 & 0.970 & 0.930 & 0.941 & 0.999 & 0.641 \\
PyTorch MLP   & 0.961 & 0.962 & 0.975 & 0.974 & 0.931 & 0.942 & 0.999 & 0.914 \\
\bottomrule
\end{tabular}}
\end{table}

\begin{table}[H]
\centering
\caption{Mean accuracy and macro F1 for Fashion MNIST, CIFAR 10, HIGGS,
high dimensional parity, high rank noise, and Covertype.}
\label{tab:ext2}
\resizebox{\textwidth}{!}{%
\begin{tabular}{lrrrrrrrrrrrr}
\toprule
 & \multicolumn{2}{c}{Fashion} & \multicolumn{2}{c}{CIFAR 10}
 & \multicolumn{2}{c}{HIGGS} & \multicolumn{2}{c}{Parity}
 & \multicolumn{2}{c}{High rank noise} & \multicolumn{2}{c}{Covertype} \\
\cmidrule(lr){2-3}\cmidrule(lr){4-5}\cmidrule(lr){6-7}
\cmidrule(lr){8-9}\cmidrule(lr){10-11}\cmidrule(lr){12-13}
Method & Acc & F1 & Acc & F1 & Acc & F1 & Acc & F1 & Acc & F1 & Acc & F1 \\
\midrule
Amplitude   & 0.821 & 0.818 & 0.378 & 0.370 & 0.643 & 0.637 & 0.485 & 0.484 & 0.795 & 0.795 & 0.587 & 0.213 \\
Angle       & 0.823 & 0.824 & 0.306 & 0.307 & 0.658 & 0.650 & 0.497 & 0.496 & 0.780 & 0.780 & 0.725 & 0.537 \\
Basis       & 0.819 & 0.819 & 0.328 & 0.326 & 0.669 & 0.665 & 0.504 & 0.504 & 0.699 & 0.699 & 0.724 & 0.555 \\
\midrule
Raw linear  & 0.797 & 0.798 & 0.276 & 0.275 & 0.641 & 0.634 & 0.486 & 0.485 & 0.791 & 0.791 & 0.720 & 0.522 \\
RFF         & 0.805 & 0.803 & 0.362 & 0.357 & 0.588 & 0.576 & 0.496 & 0.495 & 0.696 & 0.696 & 0.712 & 0.503 \\
PCA         & 0.831 & 0.830 & 0.367 & 0.364 & 0.614 & 0.606 & 0.488 & 0.486 & 0.792 & 0.792 & 0.671 & 0.313 \\
Poly 2      & n/a   & n/a   & n/a   & n/a   & 0.679 & 0.673 & 0.497 & 0.497 & 0.649 & 0.648 & 0.765 & 0.649 \\
MLP         & 0.854 & 0.853 & 0.413 & 0.413 & 0.747 & 0.745 & 0.500 & 0.496 & 0.784 & 0.784 & 0.859 & 0.779 \\
RBF SVM     & 0.853 & 0.851 & 0.457 & 0.454 & 0.679 & 0.674 & 0.492 & 0.492 & 0.795 & 0.795 & 0.746 & 0.493 \\
PyTorch MLP & 0.859 & 0.859 & 0.432 & 0.432 & 0.748 & 0.747 & 0.524 & 0.524 & 0.777 & 0.777 & 0.856 & 0.780 \\
\bottomrule
\end{tabular}}
\end{table}

\subsection{Pairwise CKA Among QIE Representations}
\label{app:cka}

Table~\ref{tab:cka_full} reports pairwise CKA values between the three QIE
representations. These values clarify that the encodings are not
interchangeable. Amplitude and angle encoding can be closely aligned on some
synthetic and physics tasks, while basis encoding is often much more distinct,
especially on image datasets.

\begin{table}[H]
\centering
\caption{Pairwise CKA between QIE encodings. Values near one indicate similar
Gram matrix structure; values near zero indicate geometrically distinct
representations.}
\label{tab:cka_full}
\small
\begin{tabular}{lrrr}
\toprule
Dataset & CKA(Amp, Angle) & CKA(Amp, Basis) & CKA(Angle, Basis) \\
\midrule
Wine            & 0.317 & 0.178 & 0.433 \\
Breast Cancer   & 0.561 & 0.249 & 0.434 \\
Dry Bean        & 0.582 & 0.288 & 0.448 \\
Credit Fraud    & 0.166 & 0.145 & 0.189 \\
Fashion MNIST   & 0.740 & 0.004 & 0.011 \\
CIFAR 10        & 0.629 & 0.001 & 0.001 \\
HIGGS           & 0.914 & 0.581 & 0.685 \\
High dimensional parity & 0.962 & 0.359 & 0.358 \\
High rank noise & 0.983 & 0.177 & 0.190 \\
Covertype       & 0.291 & 0.175 & 0.887 \\
\bottomrule
\end{tabular}
\end{table}

\end{document}